\documentclass[prb,preprint,superscriptaddress,nobibnotes]{revtex4}
\usepackage{hyperref}
\usepackage[dvips]{epsfig}

\usepackage{amsfonts}
\usepackage{amsmath}
\usepackage{amssymb}
\usepackage{pslatex}
\usepackage{color}
\usepackage{hhline}
\usepackage{threeparttable}
\usepackage{supertabular}
\usepackage{multirow}
\usepackage{array}
\usepackage{tabularx}

\begin{document}

\title{Optomechanical zipper cavity lasers: theoretical analysis of tuning range and stability}

\author{Thiago P. Mayer Alegre~$^\dag$, Raviv Perahia~$^\dag$, Oskar Painter~$^\star$}
%\address{Thomas J. Watson, Sr., Laboratory of Applied Physics, California Institute of Technology, Pasadena, CA 91125}
\affiliation{Thomas J. Watson, Sr., Laboratory of Applied Physics, California Institute of Technology, Pasadena, CA 91125}
\email{$^\star$opainter@caltech.edu}
\homepage{http://copilot.caltech.edu}

\begin{abstract}
The design of highly wavelength tunable semiconductor laser structures is presented. The system is based on a one dimensional photonic crystal cavity consisting of two patterned, doubly-clamped nanobeams, otherwise known as a ``zipper'' cavity.  Zipper cavities are highly dispersive with respect to the gap between nanobeams in which extremely strong radiation pressure forces exist.  Schemes for controlling the zipper cavity wavelength both optically and electrically are presented.  Tuning ranges as high as $75$~nm are achieved for a nominal design wavelength of $\lambda=1.3~\mu$m.  Sensitivity of the mechanically compliant laser structure to thermal noise is considered, and it is found that dynamic back-action of radiation pressure in the form of an optical or electrical spring can be used to stabilize the laser frequency.  Fabrication of zipper cavity laser structures in GaAs material with embedded self-assembled InAs quantum dots is presented, along with measurements of photoluminescence spectroscopy of the zipper cavity modes.
\end{abstract}

%\ocis{(220.4880)~Optical design and fabrication: Optomechanics, (230.4685)~Optical devices: Optical microelectromechanical devices, (250.5960)~Semiconductor lasers}

%\bibliography{./reference_zipper}
%\bibliographystyle{apsrev}
%\bibliographystyle{./ieeetr}
%\bibliography{./reference_zipper}

\maketitle

\section{Introduction}
\label{sec:Intro}

The tuning functionality of semiconductor laser cavities is an important attribute for a variety of applications, from spectroscopy~\cite{hard_laser_1970, liu_novel_1981, duarte_tunable_2008} and lightwave communication~\cite{ref:Agrawal}, to cavity quantum-electro-dynamics (QED) studies of light matter interactions~\cite{srinivasan_linear_2007}.  In this context, there has been great interest in extending both the tuning bandwidth and the tuning speed~\cite{ref:Bruce_E, liu_review_2007}.  Injection-current tuning of semiconductor lasers~\cite{coldren_continuously-tunable_1987}, through the refractive index dependence of the semiconductor on carrier concentration, can provide large bandwidth (GHz and beyond~\cite{y._matsui_30-ghz_1997}) but has limited tuning range.  Temperature tuning is also widely used for frequency stabilizing and trimming of lasers~\cite{kitamura_high-power_1983}, but is limited in both range and speed. Novel optofluidic techniques have been demonstrated in semiconductor lasers in the mid-IR~\cite{ref:Diehl, ref:Moreau_V3}, however, such techniques are again typically slow and involve integration of a somewhat cumbersome fluidic delivery system.  Faster, more dramatic tuning mechanisms have been achieved through micro-electro-mechanical-systems (MEMS). For example, tuning of a mirror on top of vertical cavity surface emitting lasers~\cite{ref:Huang_MCY}, tuning  of passive microdisks via  MEMS actuators~\cite{ref:Lee_MCM}, and electromechanical tuning of microdisks~\cite{ref:Bowen_WP}.  More recently, there has been increased interest in cavity optomechanical systems, in which optical forces through various forms of radiation pressure are used to mechanically manipulate the optical cavity~\cite{kippenberg_cavity_2008}.  Such systems have the dual property of being highly dispersive versus mechanical actuation, and can be implemented in guided wave nanostructures.  In this article, we consider the properties of a photonic crystal (PC) ``zipper'' optomechanical cavity~\cite{chan_optical_2009, eichenfield_picogram-_2009} as it pertains to on-chip semiconductor lasers.  In particular, we propose a master-slave cavity system for optical-force actuation of the zipper cavity, and explore the range of tuning and frequency stability in such mechanically compliant laser cavity structures.  We also compare optical force tuning to more conventional electrostatic tuning methods in such cavity structures.  Finally, initial photoluminescence measurements of fabricated zipper cavities in GaAs membranes containing an active region consisting of self-assembled InAs quantum dots is presented.

\begin{figure}
\begin{center}
\includegraphics[width=0.95\textwidth]{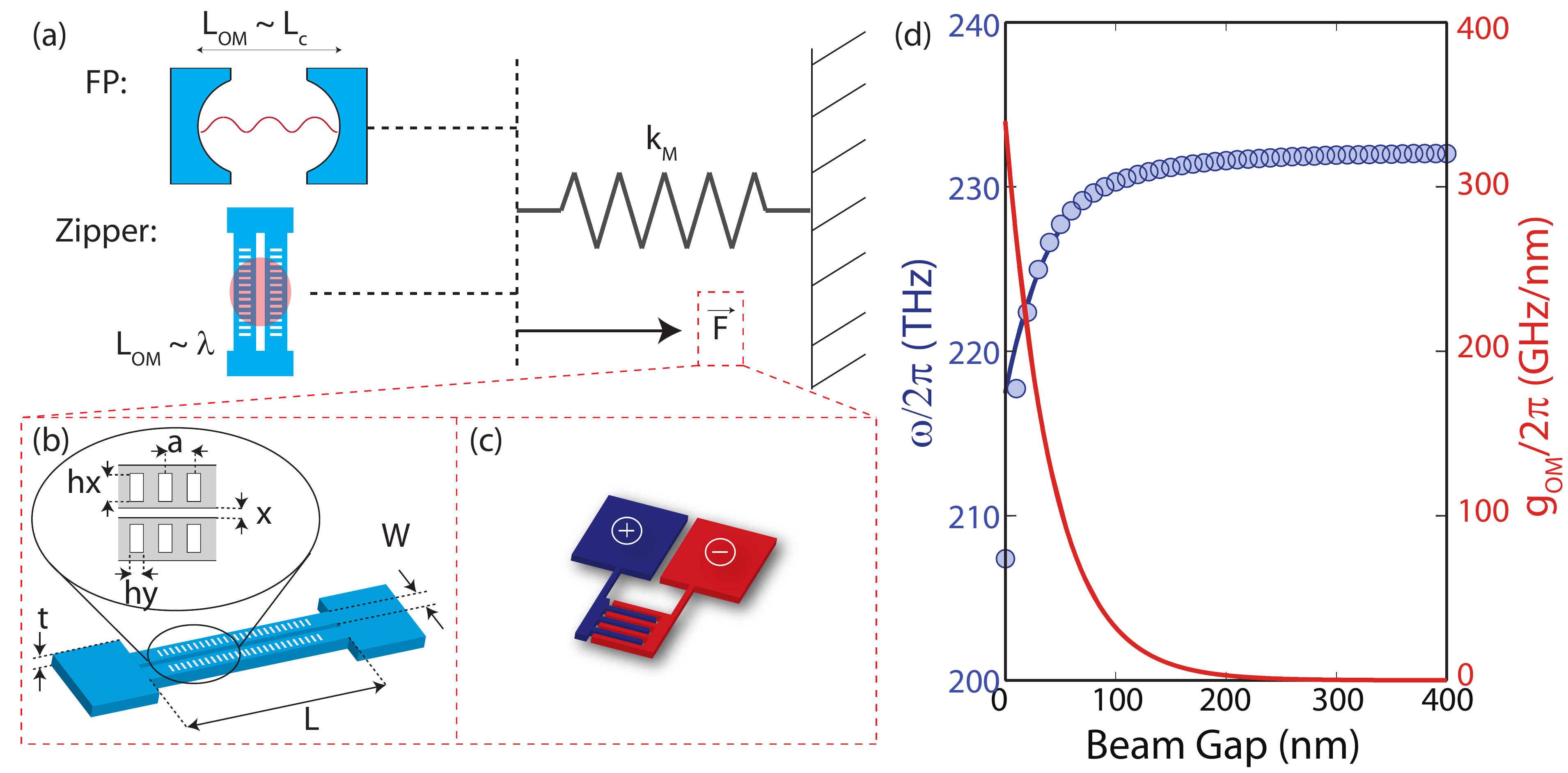}
\caption{(a) Master-slave scheme for tuning of an optomechanical cavity. (b) Master zipper cavity. (c) Master capacitive MEMS actuator. (d) Frequency dispersion and optomechanical coupling coefficient ($g_{\text{OM}}$) of a zipper cavity designed for $\lambda=1310$~nm operation as a function of gap size.}
\label{fig_intro}
\end{center}
\end{figure}

The zipper cavity structure considered in this work is shown in schematic form in Fig.~\ref{fig_intro}(b).  The zipper cavity is comprised of two doubly-clamped nanobeams in which a one dimensional PC pattern has been applied to the beams.  Such an optical cavity is highly sensitive to the gap between the two nanobeams, resulting in extremely large dispersion of the cavity frequency with nanobeam gap and a corresponding large radiation pressure force (per intra-cavity photon) pulling the beams together. As an illustration, the optical frequency dispersion as a function of gap for a GaAs zipper cavity designed for operation at $\lambda=1310$~nm ($W=400$~nm (width), $t=250$~nm (thickness), $\ell=15~\mu$m (total length)) is simulated via finite-element-method (FEM) and plotted in Fig.~\ref{fig_intro}(c).  The strength of the radiation pressure force can be quantified by the derivative of cavity frequency ($\omega_c$) versus nanobeam gap ($x_{nb}$), $g_{\text{OM}}=d\omega_c/dx_{nb}$, and is shown to exponentially increase with decreasing nanobeam gap. The large optomechanical coupling provided by zipper cavity structures clearly makes them very good candidates for tunable laser cavities.  Along with their extremely small mass, on the order of picograms, the optomechanical coupling strength of zipper cavities is on the order of $\omega_c/\lambda_c$, equivalent to the exchange of photon momentum with the mechanical system every optical cycle.  Fabry P\'{e}rot (FP) cavity systems (Fig.~\ref{fig_intro}(a)) typically have a mass on the order of grams and an optomechanical strength that scales with the inverse of the round trip time of the cavity.  For comparison an effective length scale may be defined ($L_{\text{OM}}$) over which photon momentum is transferred from the light field to the mechanical system.  In the case of a FP cavity $L_{\text{OM}}$ is equal to the physical cavity length ($L_c$),  whereas in the zipper cavity $L_{\text{OM}} \sim \lambda_{c}$.

In order to provide wideband, stable, external tuning of the zipper laser cavity, in this work we propose a scheme in which a master actuator is \emph{mechanically} coupled to a slave laser cavity. In this scheme, such a master actuator can either be another zipper cavity optical mode (Fig.~\ref{fig_intro}(b)), an entirely different zipper cavity, or a capacitively actuated MEMS structure (Fig.~\ref{fig_intro}(c)).  The mechanical stiffness of the slave cavity can be tailored and is linked to both the tunability of the structure as well as its susceptibility to thermal, and consequently, frequency noise. Since the master actuator is mechanically attached to the slave cavity, upon evaluating the tuning range we have to take into account the effective spring constant of the total system. On the other hand the frequency noise is closely related to the effective spring constant seen by small movements around a stable position, which can as well be affected by the master actuator. Here we consider and evaluate both phenomena.

This work will be divided into three main sections. In Sec.~\ref{sec:Opto_mech} we explore an all-optical master-slave zipper cavity system where different order optical cavity modes of the same structure are used.  In Sec.~\ref{sec:Electro_opto} a capacitive MEMS actuator (master) is used to control the (slave) zipper cavity. An experimental demonstration of an active quantum dot based GaAs zipper cavity will be presented in Sec.~\ref{sec:QD_zipper}. Finally, parallels and differences between optical and electrical control will be highlighted, including discussion of noise control using the optical spring effect and its electrical counterpart.

\section{All-optical optomechanical system}
\label{sec:Opto_mech}

\begin{figure}
\begin{center}
\includegraphics[width=0.95\textwidth]{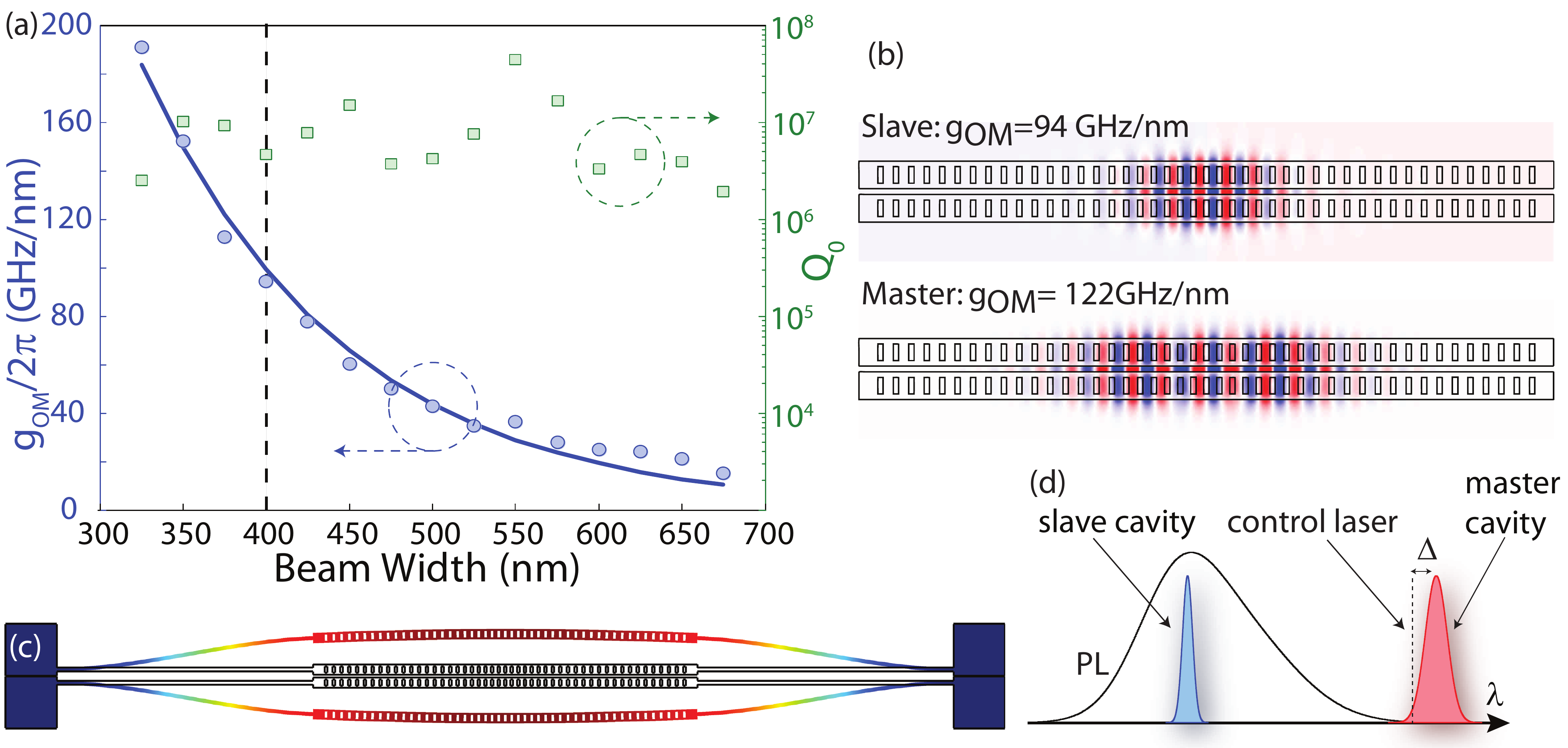}
\caption{(a) $Q$ calculated via FEM simulations and optomechanical coupling ($g_{\text{OM}}$) calculated via FEM simulations and perturbation theory are plotted as a function of beam width ($W$) for the slave cavity at $\lambda=1310$~nm.  (b) FEM simulated electric field component ($E_\text{y}$) for the $1$st (slave) and $3$rd (master) order modes at $\lambda=1310$~nm and $\lambda=1388$~nm respectively. (c) Mechanical FEM simulation of inplane motion coupling to the optical modes ($f_{\text{mech}}=600$~kHz, $m_{\text{eff}}=7.15$~pg, $k_{\text{M}}=0.1$~N/m). (d) Schematic of master-slave all optomechanical cavity system in the context of material with gain.}
\label{fig_optical_tuning_a}
\end{center}
\end{figure}

We begin by designing a single cavity all optical master-slave zipper tuning system shown in Fig.~\ref{fig_intro}(a-b). The PC defect cavity is designed in such a way that the first few orders maintain comparable quality factors ($Q$-factors).  Optical modes of a zipper cavity come in pairs of bonded and anti-bonded modes which are associated with the parity of the super mode, created by the interaction of the modes of the two nanobeams, across the gap.  A detailed design of zipper cavities can be found in Ref. \cite{chan_optical_2009}. As the optical modes are orthogonal, one can view the system as having several independent optical modes that are mechanically coupled. One can then choose the fundamental mode with frequency $\omega_{\text{s}}$ as the ``slave'' to be controlled by a higher order ``master'' mode with frequency $\omega_{\text{m}}$ (Fig.~\ref{fig_optical_tuning_a}(d)).  The tuning of the slave mode ($\Delta\omega_{\text{s}}$) is related to the nanobeam gap ($x_{nb}$) by the dispersion relation $\Delta\omega_{\text{s}}= g_{\text{OM}}^{({\text{s}})}\Delta x_{nb}$.  When power is dropped ($P_{d}$) into the master mode, a net optical force proportional to the number of photons ($N_{\text{m}}$) and the per-photon force $\left(\hbar g_{\text{OM}}^{(\text{m})}\right)$ is applied to the clamped nanobeam system with mass $m_{\text{eff}}$ and spring constant $k_{\text{M}}$, bringing the nanobeams closer together for the bonded mode. The effective mass is the motional mass of the structure and the spring constant is related to the mechanical resonance frequency $\Omega_{\text{M}}=\sqrt{k_{\text{M}}/m_{\text{eff}}}$. Since the master and slave cavities are coupled by the same nanobeam gap, the total tuning experienced by the slave cavity is given by:

\begin{eqnarray}
\label{eq_tuning}
\Delta\omega_s= g_{\text{OM}}^{(\text{s})}\frac{Q_o^{(\text{m})}g_{\text{OM}}^{(\text{m})}}{k_{\text{M}}\omega_{\text{m}}^2}P_{d}.
\end{eqnarray}

\noindent As one can see from Eq.~(\ref{eq_tuning}), large tuning relies on optimizing the opto-mechanical coupling of both modes and minimizing $k_{\text{M}}$. While reducing $k_{\text{M}}$ one must take care not to reduce the mechanical frequency below the intended frequency response of the tunable system.  Zipper cavities typically have a motional mass around tens of picograms and optomechanical coupling of $10-100$~GHz/nm yielding large tuning ranges.  Additionally, the zipper cavity geometry is amenable to trimming of $\Omega_{\text{m}}$ and $k_{\text{M}}$ by adjusting the flexibility of the supporting structure. An example of a floppy structure is shown in Fig.~\ref{fig_optical_tuning_a}(c) where the addition of struts~\cite{camacho_characterization_2009} reduced the spring constant by an order of magnitude to $k_{\text{M}}\approx0.1$~N/m for the inplane motion.

To optimize $g_{\text{OM}}$ for semiconductor zipper cavities we begin by gaining intuition from perturbation theory of moving boundary conditions~\cite{johnson_perturbation_2002,eichenfield_optomechanical_2009}:
\begin{eqnarray} \label{eq_gOM_perturb}
g_{\text{OM}} = \frac{\omega_0}{2}\frac{
\int dA \Big{(}\Delta\varepsilon|\mathbf{E}_{\parallel}|^2 + \Delta(\varepsilon^{-1})|\mathbf{D}_{\perp}|^2
\Big{)}}
{\int dV|\mathbf{E}|^2}
\end{eqnarray}

\noindent where $\Delta\varepsilon=\varepsilon_1-\varepsilon_2$, $\Delta(\varepsilon^{-1})=\varepsilon_1^{-1}-1/\varepsilon_2^{-1}$, $\varepsilon_1$ is the dielectric constant of the cavity, $\varepsilon_2$ is the dielectric constant of the surrounding, and the integration takes place over the faces perpendicular to the displacement vector.  One can therefore increase $g_{\text{OM}}$ by increasing the electric field at the boundaries normal to the direction of motion. Since the fundamental motion of these structures are inplane, i.e. the motion changes only the zipper gap size, an increase of the optomechanical coupling can be achieved by reducing both the width and gap of the waveguide beams, squeezing the field out towards the boundaries, without compromising the radiation limited $Q$. Initial fabrication of zipper cavities in $255$~nm thick GaAs and InP membranes indicated that a gap of $x_{nb}=80$~nm is a reasonable gap to achieve. Optical FEM simulations of the first order (slave) bonded defect mode (Fig.~\ref{fig_optical_tuning_a}(b))  at $\lambda=1310$~nm are carried out as a function of beam width ($W=300-675$~nm). The nanobeams are $t=255$~nm thick with an approximate refractive index $n=3.4$.  The PC mirror section of the cavity has a lattice constant $a/\lambda=0.254$ and fill fraction ($h_y/a=0.35$, $h_x/W=0.60$). The parabolic defect region is $15$ holes wide with a maximum lattice constant reduction of $15\%$. The  $Q$ and $g_{\text{OM}}$, calculated via the perturbation theory of Eq.~(\ref{eq_gOM_perturb}), are plotted as a function of $W$ in Fig. \ref{fig_optical_tuning_a}(a). We choose a beam width $W=400$~nm that maintains a high radiation limited $Q^{(\text{s})}\approx 10^7$ and high optomechanical coupling $g_{\text{OM}}^{(s)}=94$~GHz/nm. For this geometry we calculate a $g_{\text{OM}}^{(\text{m})}=122$~GHz/nm and $Q^{(\text{m})}\approx 5\times10^6$ for the $3$rd order master mode at $\lambda=1388$~nm (Fig.~\ref{fig_optical_tuning_a}(b)).

\begin{figure}
\begin{center}
\includegraphics[width=0.95\textwidth]{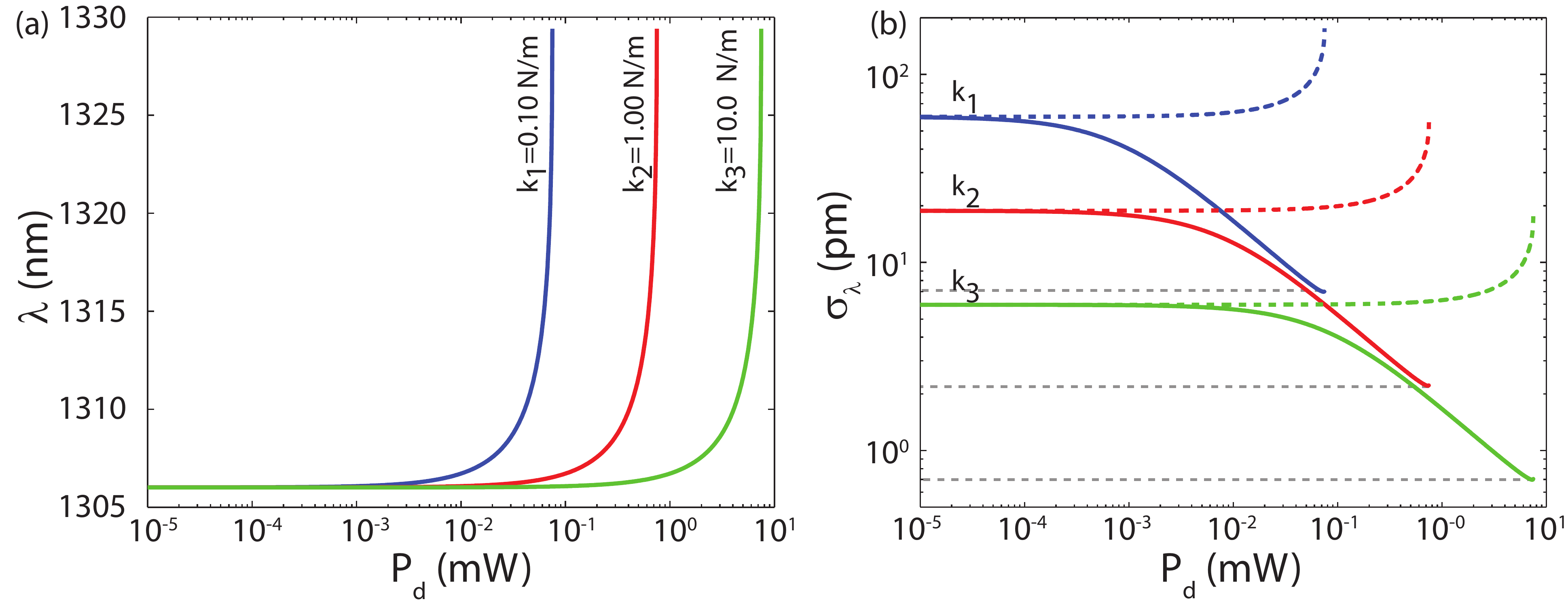}
\caption{(a) Slave mode tuning as a function of dropped pump power ($P_{d}$) into the master cavity mode for three distinct mechanical spring constants ($k=0.1,1.0$ and $10$~N/m). (b) Standard deviation of wavelength ($\sigma_\lambda$) due to thermal fluctuations at $T=300$~K (room temperature) of the slave mode as a function of $P_\text{d}$ into the master mode. Solid (dashed) lines are for $\Gamma_{t}/2$ (zero)  detuning from the master cavity resonance. The gray dashed lines indicate the minimum wavelength deviation before the nanobeams stiction point.}
\label{fig_optical_tuning_b}
\end{center}
\end{figure}

Optical FEM simulations are then used to generate a cavity dispersion curve, for both modes, as a function of gap.  A fit of the dispersion curves to a simple exponential model $\omega(x_{nb})=(\omega_0-\omega_f)e^{-\alpha x_{nb}}+\omega_f$ is then used to describe the gap-dependence of the optomechanical coupling.  In this formula $\alpha$ is a decay constant describing the exponential sensitivity of the light field in the gap, $\omega_0$ is the cavity frequency when the beams are touching, and $\omega_f$ is the cavity frequency when the beams are far apart and do not interact. Optomechanical coupling can then be modeled as $g_{\text{OM}}(x_{nb})=g_{\text{OM}}(0)e^{-\alpha x_{nb}}$. A plot of $\omega(x_{nb})$, the fit, and $g_{\text{OM}}(x_{nb})$ for the slave mode is plotted in Fig.~\ref{fig_intro}(d). From this, the slave mode tuning can now be calculated as a function of the dropped optical power into the master mode, $P_{d}$, using Eq.~\ref{eq_tuning}. As the mechanical spring constant of zipper cavities can be tailored, we choose several representative values ($k_\text{M}=0.1$, $1.0$, and $10.0$~N/m) to simulate throughout this work which encompass experimentally feasible systems. Tuning of the slave mode versus master $P_{d}$ is calculated for three mechanical spring constants and shown in Fig.~\ref{fig_optical_tuning_b}(a).  Since the applied optical force, for a constant dropped optical power, is nonlinear (exponential) with nanobeam gap, there is a position at which any additional optical force cannot be made up for by the restoring mechanical spring force, and no stable equilibrium can be achieved without the nanobeams touching. This position, for the minimum allowed slot gap ($x_{\text{nb,min}}$), is known in the MEMS community as the \emph{in-use stiction} point~\cite{MEMS_tas_stiction_1996, MEMS_maboudian_critical_1997}.
An analytical expression for minimum gap before stiction can be found, $x_\text{nb,min}=x_\text{nb,0}-\alpha^{-1}$, where $x_\text{nb,0}$ is the fabricated (initial) nanobeam gap, as long as the optomechanical coupling has the above simple exponential behavior and one assumes the dropped optical power is independent of gap. Under such circumstances, a maximum allowed dropped power for the master mode is equal to $P_{d}^{\text{max}} = k_{\text{M}}\omega_{\text{m}}^2/(2Q_o^{(m)}g_{\text{OM}}^{(m)}(0)\alpha e^{-\alpha x_\text{nb,min}})$, resulting in a maximum tuning range of the slave mode for the geometry considered here with $k_\text{M}=0.1$~N/m of $\Delta\lambda=25$~nm.  For this spring constant the maximum power required to obtain the full tuning range is only $P_{d}=58~\mu$W. In some experimental realizations a more nuanced analysis of the stiction point is necessary. For example, when a pump laser is used, an increase in pump power leads to cavity laser detuning. The maximum tuning range presented herein is therefore a conservative estimate and applies to situations where a wide band pump source such as a diode can be used.

Obviously, less stiff structures allow for lower power actuation; however, along with smaller spring constant comes a susceptibility to thermal Brownian motion. We estimate the standard deviation of the resonant wavelength ($\sigma_{\lambda}$) of the slave laser cavity by relating the position variance to the temperature and spring constant~\cite{rosenberg_static_2009}.  As can be seen from Fig.~\ref{fig_optical_tuning_b}(b), for a $k_\text{M}=0.1$~N/m device (dashed line) the wavelength noise will increase from $\sigma_{\lambda}=60$~pm to $\sigma_{\lambda}=100$~pm by the time its tuned the full range. The thermal Brownian motion can be significantly decreased using the optical spring effect~\cite{kippenberg_cavity_2007, braginsky_quantum_1995} of the master mode. In the sideband unresolved regime, the in-phase component of the optical force leads to a modified effective dynamic spring constant~\cite{eichenfield_picogram-_2009, rosenberg_static_2009,lin_mechanical_2009}:

\begin{eqnarray}
\label{eq_opt_spring}
k_\text{M}'=k_\text{M}+\frac{2g_{\text{OM}}^2P_{d}\Delta}{\omega_c\Gamma_t{[\Delta^2+(\Gamma_t/2)^2]}},
\end{eqnarray}

\noindent where $\Delta=\omega_l-\omega_c$ is the detuning of the input laser ($\omega_l$) from the cavity resonance ($\omega_c$) and $\Gamma_t$ is the loaded optical cavity loss rate. The minimum frequency noise occurs for a maximum optical spring effect achieved when $\Delta=\Gamma_t/2$.  A comparison of wavelength noise with and without the optical spring effect for optimal detuning is shown on Fig.~\ref{fig_optical_tuning_b}(b).  At maximum tuning, for $k_\text{M}=0.1$~N/m, the increase in the dynamic spring constant decreases the thermal noise from $\sim100$~pm to $\sim9$~pm.

Despite the fact that the thermal noise is initially higher for lower spring constant at zero detuning, the optical spring effect can compensate for the noise and allow tuning of the slave cavity through the full range with one hundredth of the power required for the higher spring constant. From a practical standpoint, a residual broadening of the laser line due to thermal noise of tens of picometers will not significantly degrade the laser linewidth in the case of semiconductor nanolasers.  Such small-volume lasers typically have linewidths on the order of $\delta\lambda=100$-$200$~pm \cite{ref:Perahia_R,ref:Hill_MT}, limited by the small system size of the laser and the resulting large fractional fluctuations in the carrier density, spectral-hole burning, and related thermal effects~\cite{ref:Carmichael_Rice, ref:Yamamoto_Bjork4, slusher_optical_1994}.

\section{Electro-optomechanical system}
\label{sec:Electro_opto}

\begin{figure}
\begin{center}
\includegraphics[width=0.855\textwidth]{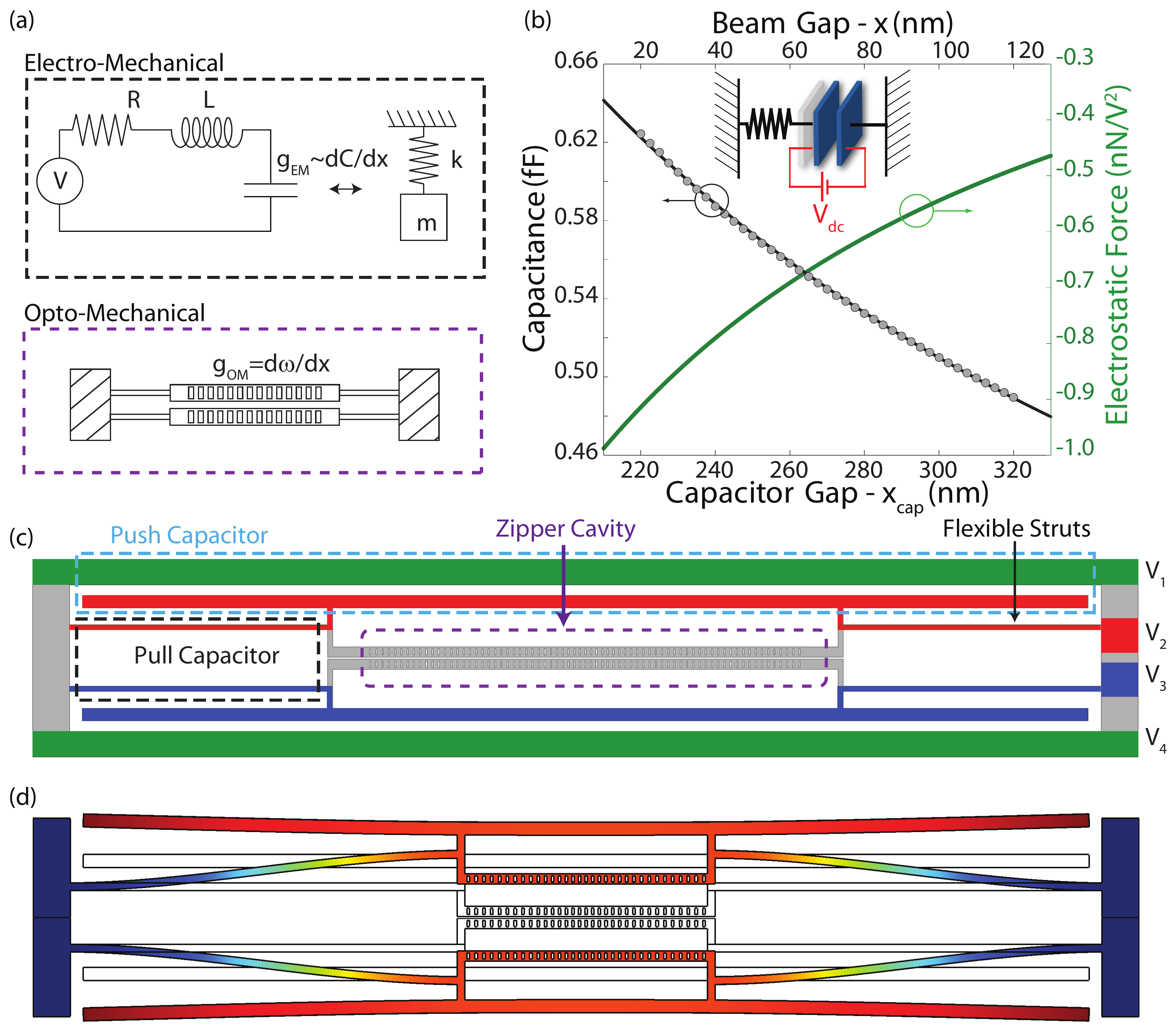}
\caption{(a) The MEMS actuation system can be described as a RLC circuit coupled to a spring-mass system, where the coupling coefficient $g_\text{EM}$ is proportional to the change in the MEMS capacitance. This coupling strength can be describe in the same way of optical system, $g_\text{EM} = d\Omega_\text{LC}/dx_\text{cap}$. (b) FEM simulation of the capacitance for the MEMS design showed in (c) as a function of the capacitor gap ($x_\text{cap}$). On the right the evaluated force per applied voltage is also shown. The negative sign accounts for the attractive source of the force. (d) The two mechanical motion created by the push and pull capacitor.}
\label{fig_electro_mechanical}
\end{center}
\end{figure}

Based upon the optical design for the slave cavity presented in the previous section (Sec. \ref{sec:Opto_mech}), we develop a master actuator using electrostatic forces to provide the tuning. The MEMS design for actuating the optical cavity is shown in Fig.~\ref{fig_electro_mechanical}(c).  Four electrodes allow for both pulling the nanobeams together (pull capacitor) as well as pushing them apart (push capacitor). As opposed to using a semiconductor capacitor~\cite{loncar_dynamically_2009}, we propose using metal electrodes insulated from the active region by a dielectric layer, significantly reducing charging time below the frequency response of the mechanical structure.  Separation between the optical structure and metal capacitor structure avoids metal or doping induced optical loss as well as bias induced free-carrier optical loss.  Dynamic mechanical FEM simulations of the mechanical structures, with exaggerated displacement, are shown in Fig.~\ref{fig_electro_mechanical}(d) yielding a mechanical resonance frequency of $\Omega_\text{M}=200$~kHz. Since for wavelength tunability and stability we are interested in the spring constant as seen by the centered optical cavity, static mechanical FEM simulations are used to calculate a spring constant $k_\text{M}=0.7$~N/m at the center of the structure. From here on we analyze the pulling force exerted by the metal over the struts as the pushing force can be simulated in an analogous fashion.

%Talk about the DC tuning range
The MEMS structure can be modeled as a parallel plate capacitor with one of the plates coupled to a spring (inset on Fig.~\ref{fig_electro_mechanical}(b)).
The capacitance for the MEMS structure as a function of the gap size between struts is simulated via FEM and plotted in Fig.~\ref{fig_electro_mechanical}(b). Since the capacitance of the structure is on the order of femtofarads (fF), when the structure is experimentally implemented, an appropriately designed RLC tank circuit will be used to reduce spurious impedances (Fig.~\ref{fig_electro_mechanical}(a)~\cite{tank_sillanpaa_accessing_2009}).  For a DC voltage ($V_{dc}$), the applied attractive force on each strut as a function of the capacitor gap size ($x_\text{cap}$) can be evaluated as $F_\text{el}=(1/2)\left(dC/dx_\text{cap}\right)V_{dc}^2$. The total displacement, proportional to force, will produce a change in the optical frequency of the slave cavity through the change in capacitance. While the displacement does not depend on resistance, it is illustrative for comparison with the all-optical case (Sec.~\ref{sec:Opto_mech}) to define the capacitive-mechanical coupling as $g_\text{CM} = d\Gamma_\text{RC}/dx_\text{cap}$, where $\Gamma_\text{RC}=1/RC$ is the capacitor charging time constant. The tuning of the slave cavity is then given by:

\begin{eqnarray}\label{eq_DC_tuning}
\Delta\omega_\text{s}= g_{\text{OM}}^{(\text{s})}\frac{2g_\text{CM}^{(\text{m})}}{k_\text{M}\Gamma_{\text{RC}}^2}\frac{V_{dc}^2}{R},
\end{eqnarray}

From Fig.~\ref{fig_electro_mechanical}(b) we can see that $g_\text{CM}$ is non-linearly dependent upon capacitor gap. As a result, we once again have a geometry-dependent relation between the minimum gap size and the maximum allowed voltage, i.e. the capacitor pull-in voltage.  Using an empirical model for the capacitance $C=a\times x_\text{cap}^{-b} + C_0$, one can evaluate the minimal gap size as $x_\text{cap, min}=(b+1)/(b+2)x_\text{cap,0}$, where $x_\text{cap,0}$ is the initial capacitor gap size and $a$ and $b$ are positive and real parameters found by fitting the empirical model to the FEM simulations (Fig.~\ref{fig_electro_mechanical}(b)). This relation limits the maximum allowed voltage to be:

\begin{eqnarray}\label{eq_stick_cap}
V_\text{max} = \left[
\left(\frac{(b+1)^{b+1}}{(b+2)^{b+2}}\right)\left(\frac{k_\text{M}}{ab}\right)x_{\text{cap},0}^{(b+2)}
\right]^{1/2}.
\end{eqnarray}

\begin{figure}
\begin{center}
\includegraphics[width=0.95\textwidth]{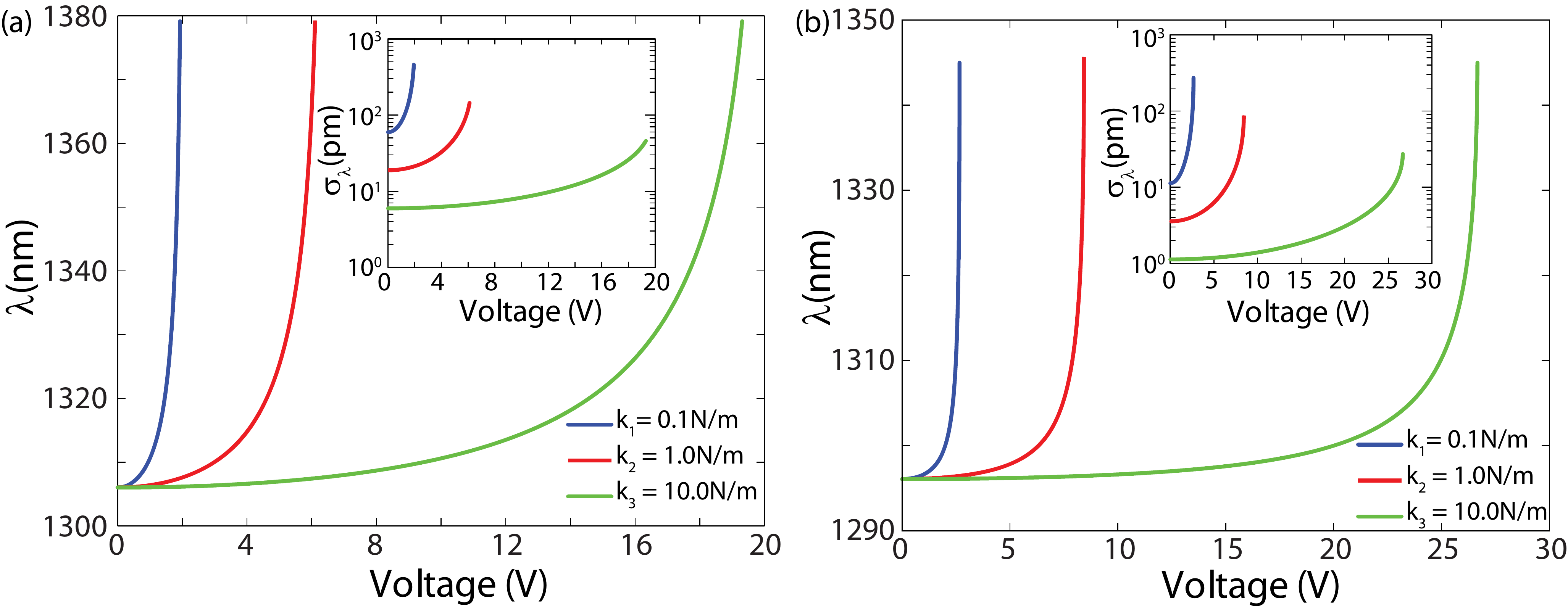}
\caption{(a) Optical tuning as a function of the applied DC voltage for spring constants between $0.1$ and $10.0$~N/m. The initial gap between beams was taken to be $80$~nm (capacitor gap $280$~nm). (b) Same as (a) but for a initial gap of $150$~nm between beams capacitor gap $350$~nm. The insets show the resulting standard deviation of the cavity wavelength due to thermal noise (room temperature) as a function of the applied voltage for both cases.}
\label{fig_DC_tuning}
\end{center}
\end{figure}

\noindent For a geometry where the initial nanobeam gap is $x_\text{nb,0}=80$~nm and the initial capacitor gap is $x_{\text{cap},0}=280$~nm the maximum displacement given by the capacitive stiction point is $\delta x_\text{nb}\sim110$~nm, clearly larger than the initial nanobeam gap. Actuation is therefore not limited by the capacitive stiction point but by the nanobeam gap. Tuning as a function of the DC voltage for three different spring constants ($k_\text{M}=0.1,1.0$ and $10$~N/m) is calculated and plotted in Fig.~\ref{fig_DC_tuning}(a).  A very large tuning range of $\Delta\lambda=75$~nm can be achieved before the nanobeams touch.  As in the all-optical case, the thermal Brownian motion at room temperature experienced by the nanobeams (Fig. \ref{fig_DC_tuning}(a) inset) results in wavelength fluctuation of the laser cavity. For the $k_\text{M}=0.1$~N/m MEMS device the wavelength noise standard deviation is $\sigma_{\lambda}=450$~pm at maximum tuning. When compared to the all optical case, the cavity can be tuned farther but is accompanied by larger noise due to the fact that there is no dynamical spring effect to reduce the thermal fluctuations of the nanobeams.

As the electrostatic actuation produces a very large displacement, one could potentially increase the initial nanobeam gap to reduce the effects of thermal Brownian motion on laser cavity linewidth.  Starting with a larger gap also eases fabrication tolerances. As a comparison, Fig.~\ref{fig_DC_tuning}(b) shows the case where the initial beam gap is $x_\text{nb, 0}=150$~nm  (capacitor gap size $x_\text{cap, 0}=350$~nm).  The total tuning range, before the capacitor stiction point, is still significant at $\Delta\lambda\sim50$~nm, but with roughly half the room temperature linewidth noise ($\sigma_{\lambda}=260$~pm) for $k_\text{M}=0.1$~N/m.

%%%%%%%%%%%%%%%%%%%%%%%%%%%%%%%%%%%%%%%%%%%%%%%%%%%%%%
%%%%%%%%%%%%%%%%%%%%%%%%%%%%%%%%%%%%%%%%%%%%%%%%%%%%%%

%AC tuning range
In an electrical analogy to the all-optical case one can apply an AC field to actuate the MEMS. By solving the equation of motion for the capacitor plate coupled to the RLC charge equation we can write for the gap change:

\begin{eqnarray}\label{eq_AC_actuation}
\Delta x = \frac{4g_\text{EM}}{k_\text{M}\Omega_\text{LC}^2Q_\text{LC}}|\eta_0|^2\times
\frac{\big{(}V_{ac}/\sqrt{2}\big{)}^2}{R},
\end{eqnarray}

\noindent where $V_{ac}$ is the amplitude of the monochromatic voltage source at frequency $\omega_0$, $g_\text{EM}=d\Omega_\text{LC}/d x_{\text{cap}}=(1/2)g_\text{CM}/Q_\text{LC}$, $Q_\text{LC}$ is the $RLC$ circuit quality factor, and the last term can be identified as the mean dropped power into the system. $|\eta_0|^2 = \Omega_\text{LC}^4/[(\Omega_\text{LC}^2-\omega_0^2)^2+\Gamma_\text{LC}^2\omega_0^2]$ is a dimensionless parameter with $\Omega_\text{LC}/2\pi=1/\sqrt{LC}$, and $\Gamma_\text{LC}/2\pi = R/L$ being the RLC resonance frequency and damping rate respectively.

As expected the RLC circuit acts as a frequency band pass filter for the driving voltage source resulting in three different cases. First, for $\omega_0\ll\Omega_\text{LC}$, $|\eta_0|^2\approx1$ which results in a displacement similar to the DC case, where now the DC voltage is replaced by the time average voltage $V_{ac}/\sqrt{2}$. Second, for $\omega_0\gg\Omega_\text{LC}$, $|\eta_0|^2\approx 0$ resulting in negligible displacement.  The third case, when $\omega_0$ is close to the circuit's resonance frequency, is an analog to the all-optical case where the RLC resonant circuit acts as the master cavity. In this case $|\eta_0|^2\approx\Omega_\text{LC}^2/(4\Delta^2+\Gamma_\text{LC}^2)$, where $\Delta = \omega_0 - \Omega_\text{LC}$ is the detuning of the input voltage from the RLC resonance. The maximum change in the gap size is reached when $\Delta=0$ which leads to a frequency shift for the optical mode give by:

\begin{eqnarray}\label{eq_AC_tuning}
\Delta \omega_\text{s} = g_{\text{OM}}^{(\text{s})}\frac{4g_\text{EM}Q_\text{LC}}{k_\text{M}\Omega_\text{LC}^2}\times
\frac{\big{(}V_{ac}/\sqrt{2}\big{)}^2}{R}.
\end{eqnarray}

Comparing this equation with Eq.~\ref{eq_DC_tuning} for the DC case, the presence of the RLC circuit is explicit.  The actuation voltage is now inversely proportional to the quality factor of the circuit.

\begin{figure}
\begin{center}
\includegraphics[width=0.95\textwidth]{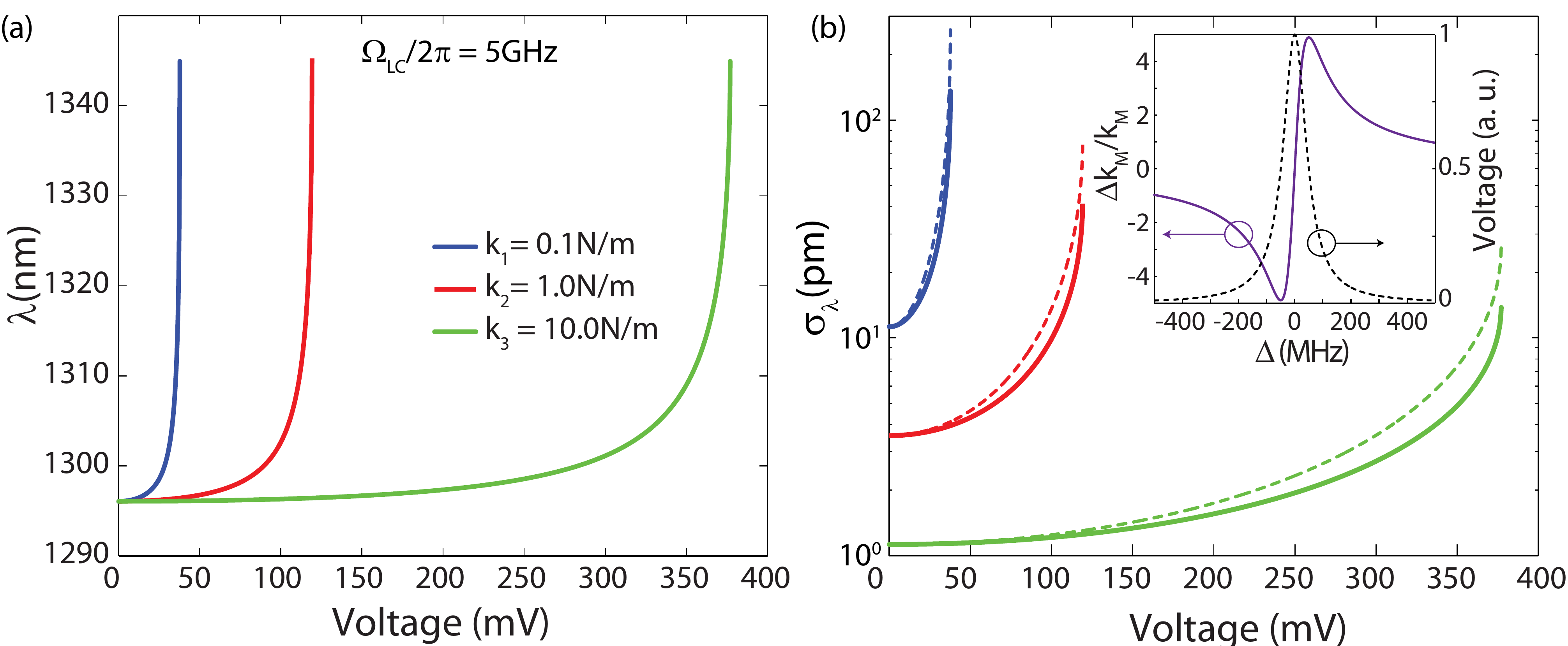}
\caption{(a) Optical tuning as a function of the applied AC voltage for spring constants between $0.1$ and $10.0$~N/m for an AC signal at $\omega_{ac}/2\pi=5$~GHz. The RLC circuit was assumed to have a quality factor of $Q_\text{LC}=100$. (b) The solid (dashed) lines shows the change in the mechanical spring constant $\Delta k_\text{M} = k_\text{M}(V=V_{ac}) - k_\text{M}(V=0)$ as a function of the voltage for the optimal (no detuning) detuning from the RLC resonance frequency. The inset shows the fraction change of the spring constant $\Delta k_\text{M}/k_\text{M}$ as one detune the input voltage from the RLC resonance frequency. The dashed black line shows the resonance of the RLC circuit.}
\label{fig_electrical_AC_tuning}
\end{center}
\end{figure}

We can now evaluate the tuning curve of the optical mode as a function of the input voltage. This is shown in Fig.~\ref{fig_electrical_AC_tuning}(a) for an illustrative set of parameters, $\Omega_\text{LC}/2\pi=\omega_0/2\pi= 5$~GHz and $Q_\text{LC}=100$. Following the DC case, the initial beam gap size was taken as $x_\text{nb,0}=150$~nm (capacitor gap size of $x_{\text{cap},0}=350$~nm) which gives the same total tuning range of $\Delta\lambda\sim50$~nm. Due to the quality factor of the resonator the applied voltage needed is reduced. For the $k_\text{M}=0.1$~N/m device we need less than $40$~mV to actuate and tune the structure through the full range.

The thermal Brownian motion at room temperature coupled to the optical cavity is the same as that of the DC case. It broadens the $k_\text{M}=0.1$~N/m device linewidth to $\sigma_{\lambda}\sim260$~pm as seen in Fig. \ref{fig_electrical_AC_tuning}(b). Since Eq.~\ref{eq_AC_tuning} is essentially the same as Eq.~\ref{eq_tuning}, but for the electrical domain~\cite{braginsky_quantum_1995}, one would expect an electric phenomena similar to the optical spring.  In this case the detuned AC field modifies the mechanical spring constant. When the mechanical frequency is much smaller than the $RLC$ resonance frequency we can evaluate the change of the spring constant as a function of the detuning $\Delta$ as:
\begin{eqnarray}
\label{eq_elec_spring}
k_\text{M}'=k_\text{M}+\frac{2g_{\text{EM}}^2}{Q_\text{LC}}\Bigg{[}
\frac{\Delta}{\big{(} \Delta^2 + (\Gamma_\text{LC}/2)^2  \big{)}^2}
\Bigg{]}
\times\frac{\big{(}V_{ac}/\sqrt{2}\big{)}^2}{R}.
\end{eqnarray}

The inset on Fig.~\ref{fig_electrical_AC_tuning}(b) shows the ratio of mechanical spring constant change $\Delta k_\text{M}/k_\text{M}$ as a function of the detuning for the maximum allowed voltage (before stiction point) .  Since we are in the side band unresolved regime the change in the spring constant does not depend on its initial value.  The solid lines on Fig.~\ref{fig_electrical_AC_tuning}(b) show the standard deviation of the cavity position resulting from thermal noise at the optimal detuning. Compared to when no detuning is applied (dashed lines) we can observe a reduction of the thermally-induced laser wavelength noise from $\sigma_{\lambda}=260$~pm to $\sim130$~pm for the $k_\text{M}=0.1$~N/m device at the full range.

\section{Photoluminescence spectroscopy of zipper cavities}
\label{sec:QD_zipper}

\begin{figure}[ht!]
\begin{center}
\includegraphics[width=0.95\textwidth]{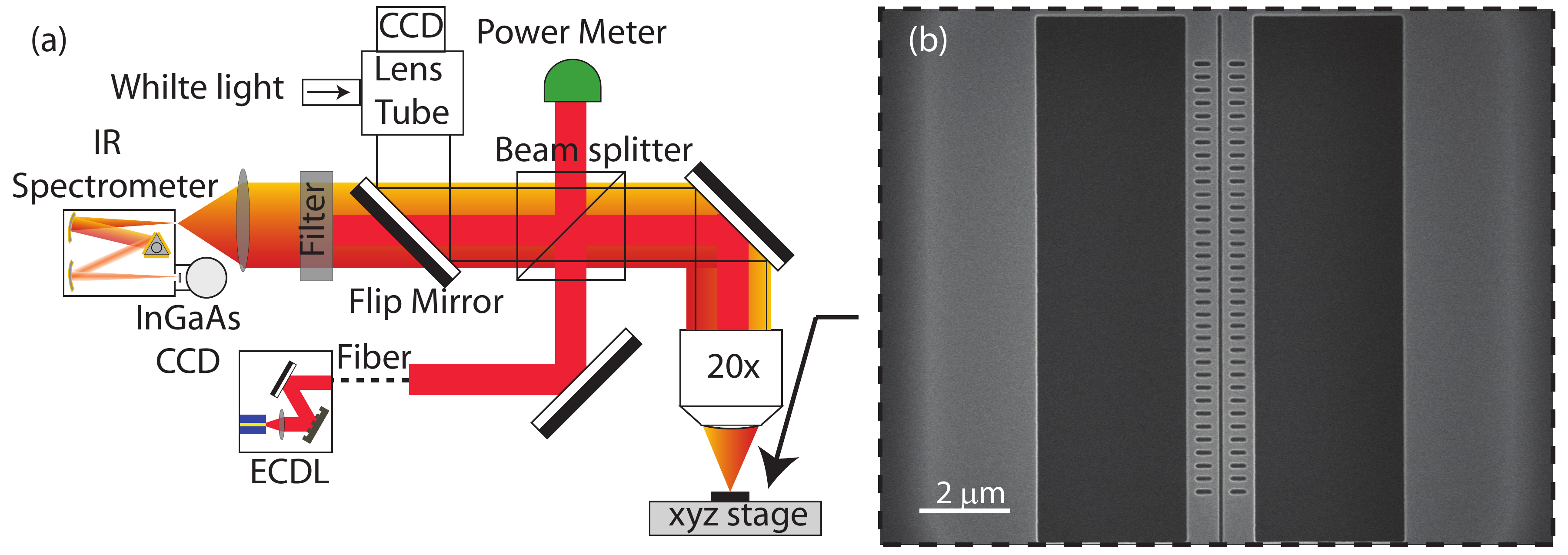}
\caption{(a) Photoluminescence experimental setup. (b) SEM micrograph top view of InAs/GaAs zipper cavity.}
\label{fig_QD_setup}
\end{center}
\end{figure}

Here we present optical mode spectroscopy of zipper optomechanical cavities formed in light-emitting GaAs material.  The active region in the GaAs material consists of embedded, self-assembled dots-in-a-well (DWELL) which emit at $\lambda\approx1.2~\mu$m~\cite{ref:Liu_G, ref:Stintz}. Zipper cavities are fabricated by first depositing a hard silicon nitride (SiN$_x$) mask ontop of a $255$~nm thick GaAs membrane layer which contains the self-assembled InAs quantum dots (QDs). The cavity is patterned via electron-beam lithography, inductively-coupled reactive-ion (ICP-RIE) etching of the SiN$_x$ mask based on C$_4$F$_8$/SF$_6$ chemistry, and subsequent pattern transfer into the QD membrane via ICP-RIE etching based on Ar/Cl$_2$ chemistry.  Devices are released from the substrate by wet etching of an underlying sacrificial AlGaAs layer in a dilute hydrofluoric acid solution ($60:1$ H$_2$O:HF)~\cite{srinivasan_cavity_2006}. Native oxide residue was removed using a dilute ammonium hydroxide solution ($25:1$ H$_2$O:NH$_4$OH)~\cite{ref:Seo_JW}.

Devices are tested in a free space photoluminescence (PL) setup shown in Fig.~\ref{fig_QD_setup}(a). The devices, SEM images of which are shown in Fig.~\ref{fig_QD_setup}(b), are free space pumped with an external cavity diode laser (ECDL) at $\lambda=830$~nm and free space photoluminescence is collected via a high numerical aperture objective and sent to a spectrometer coupled to an InGaAs charge-coupled device (CCD) array. The pump laser is pulsed to reduced thermal broadening with a period of $\tau=1~\mu$s, pulse width of $w=200$~ns, resulting in an average pump power of $P_p=317~\mu$W. A band pass filter is used to remove the residual reflected pump laser from the PL. A power meter at one port of a beam splitter measures the average incident power. The sample is mounted on top of a \emph{xyz}-stage enabling rapid testing of multiple devices.

\begin{figure}[ht!]
\begin{center}
\includegraphics[width=0.95\textwidth]{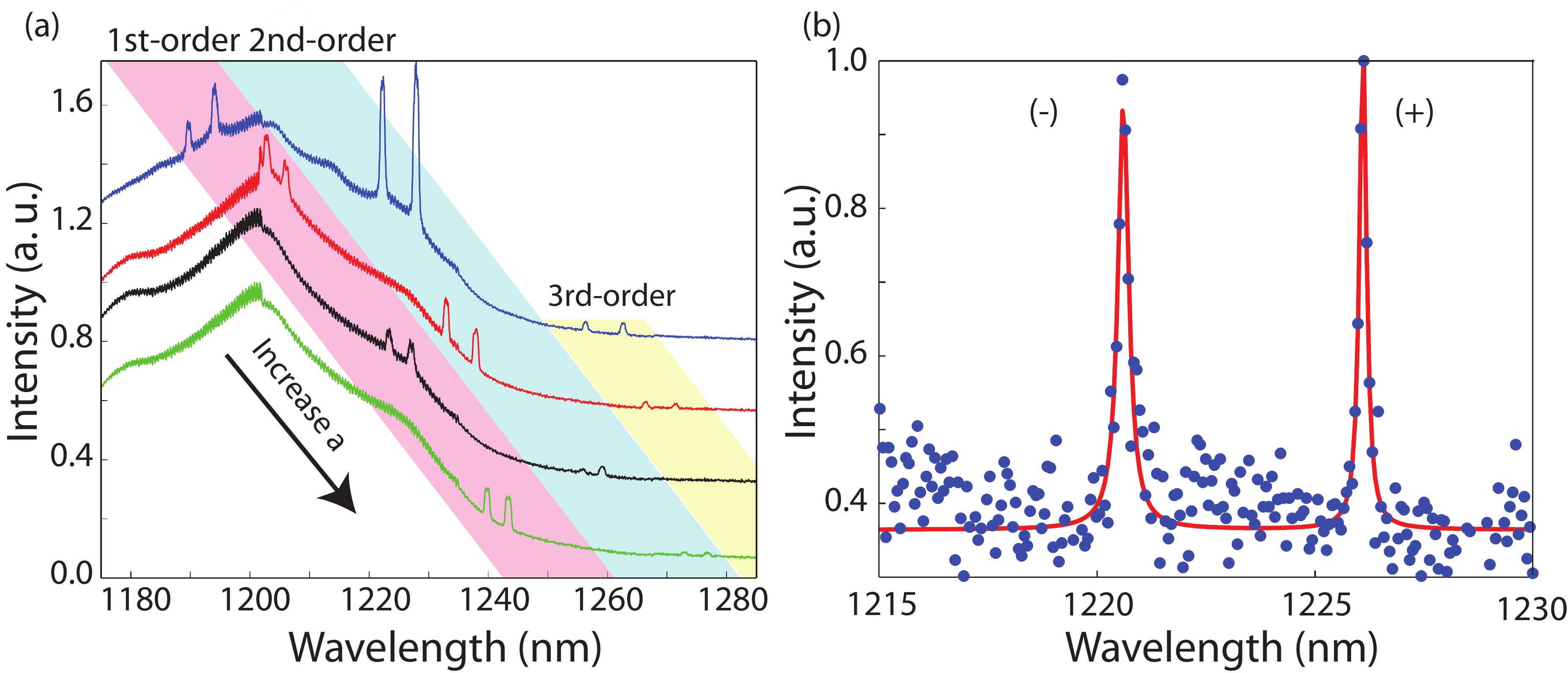}
\caption{(a)  Optically pumped zipper cavity modes spectra as a function of increasing $a/\lambda$. (b) Modes come in (+) bonded and anti bonded (-) pairs.}
\label{fig_QD_spectra}
\end{center}
\end{figure}

PL measurements show both the bonded and anti-bonded pairs as predicted and measured in passive cavities~\cite{chan_optical_2009}.  First order (red shade) through third order (yellow shade) defect cavity modes can be seen in Fig.~\ref{fig_QD_spectra}(a).  The modes collectively tune with increase of the photonic crystal lattice constant of the cavity as predicted by simulations. The wavelength splitting between bonded and anti-bonded modes is measured to be $\Delta\lambda_{\pm}=5-7$~nm.  Such a splitting indicates a nanobeam gap of roughly $100$~nm, consistent with the SEM images of the fabricated devices.

Fig.~\ref{fig_QD_spectra}(b) shows both bonded and anti-bonded modes at very low average pump power ($w=10$~ns, $\tau=4~\mu$s, $P_\text{p}=101$~nW). At this power level, and for the single layer of quantum dots in these devices, the expected enhancement/reduction to the pumped cavity $Q$-factor is negligible.  From the linewidth of the modes, we estimate the quality factor of the bonded mode in these devices to be around $Q\approx 6000$. Although not large enough to achieve lasing in this QD material (a $Q$-factor of greater than $5\times10^4$ is required to reach threshold given the limited QD gain), these results are encouraging given the much more substantial gain attainable in quantum well material\cite{ref:Hwang2}.

\section{Discussion}
\label{sec:Disc}
In the work presented here, highly dispersive zipper optomechanical cavities have been proposed and designed for tunable on-chip semiconductor lasers.  Two master-slave schemes for wavelength tuning have been proposed. The first, all-optical, allows for tuning using a higher order zipper mode. The second, electrical-optical, allows for tuning via a capacitively actuated MEMS structure integrated into the zipper cavity structure. Both schemes take advantage of the very large $g_\text{OM}$ values unique to the zipper cavity geometry. Tuning range is related to initial nanobeam separation, mechanical spring constant, the force delivered by the master, and geometric limitations. The high mechanical compliance of these structures, i.e. low static spring constant, makes them susceptible to thermal and therefore frequency noise. In some cases the susceptibility to frequency noise can be mediated by the dynamic backaction of the control optical (electrical) field and its contribution to a dynamic spring constant.

In the all optical case, tuning range may be limited by the stiction point and therefore the initial nanobeam gap. In the case of tuning with reduced noise a tunable master laser source is required.  Depending upon the geometry and application, there may also be issues associated with the absorption of the master laser light field, and its contribution to optical pumping of the slave cavity and the resulting generated free carriers. Not withstanding these potential shortcomings, we have shown that for reasonable master laser dropped power levels of  $50~\mu$W-$10$~mW, a large radiation pressure force tuning of the slave cavity can be achieved up to a range of $\Delta\lambda=25$~nm. Lower power actuation is attained by increasing the Q-factor or decreasing the static spring constant. The latter is associated with an increase in thermo-mechanical laser frequency noise.  We have shown that a self-compensation of the thermal noise occurs in the optical tuning case due to dynamical back-action of the light field on the mechanical system via the optical spring effect. Even for the lowest static spring $k_\text{M}=0.1$~N/m considered in this work, the thermo-mechanical wavelength noise is reduced to below $\sigma_{\lambda}\sim11$~pm over the full tuning range.

Analysis of a more conventional electrical-optical scheme is also performed to compare with the all-optical scheme. Two main advantages of the electro-optical scheme are the high magnitude of force available from electrostatic actuation, even at large gap distances, and the simple electrical control mechanism.  In the laser structures studied in this work, the DC electrical tuning mechanism allows for as much as $75$~nm of wavelength tuning or $5\%$ of the nominal laser frequency.  A drawback of the DC tuning mechanism is the inability to mediate thermal noise, except through an increase of the stiffness of the mechanical structure which in turn reduces the tuning range for a given applied voltage. The use of an AC voltage does allow for the increase of dynamic spring constant through electrical spring phenomena; however, the reduction of noise via the electrical spring is significantly less than in the optical case primarily because of the larger, attainable $Q$-factor in the optical domain. The $Q$ of an RLC circuit can potentially be substantially increased by moving to superconducting materials at low temperature~\cite{day_broadband_superconduting_microstrip_2003}, although going to such means removes many of the practical advantages of the electro-optical scheme.

Numerous trade-offs exist between the all-optical and electrical-optical schemes.  Although not studied here, one may envision a hybridized approach in which coarse laser frequency tuning is performed via electrostatic actuation, and fine laser frequency tuning and laser frequency stabilization are performed by an optical control field.  There are also other interesting laser phenomena in addition to simple laser tuning that may be explored in similar optomechanical laser systems.  The tuning dependence upon the master laser dropped power, and thus master laser frequency, in the all-optical case may be used as a frequency lock of master and slave lasers.  The consideration of radiation pressure effects resulting from internal radiation, as in the case of the slave laser, leads to a whole new set of nonlinear dynamical equations and effects.  Such effects include self-frequency-locking and mode-locking\cite{haus_mode-locking_2000}.  Future work will aim to theoretically and experimentally explore these new dimensions to active cavity optomechanical systems.

\section{Acknowledgments}
\label{sec:Acknowledgments}
The authors would like to thank Qiang Lin and Jessie Rosenberg for rich discussions and Jasper Chan for help with simulations. This work was supported by the DARPA NACHOS program (Award No. W911NF-07-1-0277).

$^{\dag}$These authors contributed equally to this work.

%\bibliography{./reference_zipper}

\begin{thebibliography}{10}

\bibitem{hard_laser_1970}
T.~M. Hard, ``Laser wavelength selection and output coupling by a grating,''
  {\em Appl. Opt.}, vol.~9, no.~8, pp.~1825--1830, 1970.

\bibitem{liu_novel_1981}
K.~Liu and M.~G. Littman, ``Novel geometry for single-mode scanning of tunable
  lasers,'' {\em Opt. Express}, vol.~6, pp.~117--118, Mar. 1981.

\bibitem{duarte_tunable_2008}
F.~J. Duarte, {\em Tunable Laser Applications, Second Edition}.
\newblock {CRC}, 2~ed., Aug. 2008.

\bibitem{ref:Agrawal}
G.~P. Agrawal and N.~K. Dutta, {\em {Semiconductor Lasers}}.
\newblock New York, NY: Van Nostrand Reinhold, 1993.

\bibitem{srinivasan_linear_2007}
K.~Srinivasan and O.~Painter, ``Linear and nonlinear optical spectroscopy of a
  strongly coupled microdisk-quantum dot system,'' {\em Nature}, vol.~450,
  pp.~862--865, Dec. 2007.

\bibitem{ref:Bruce_E}
E.~Bruce, ``{Tunable Lasers},'' {\em IEEE Spectrum}, vol.~39, pp.~35--39, Feb.
  2002.

\bibitem{liu_review_2007}
A.~Q. Liu and X.~M. Zhang, ``A review of {MEMS} external-cavity tunable
  lasers,'' {\em J. Micromech. Microeng.}, vol.~17, no.~1, pp.~{R1--R13}, 2007.

\bibitem{coldren_continuously-tunable_1987}
L.~Coldren and S.~Corzine, ``Continuously-tunable single-frequency
  semiconductor lasers,'' {\em IEEE J. Quantum Electron.}, vol.~23, no.~6,
  pp.~903--908, 1987.

\bibitem{y._matsui_30-ghz_1997}
Y.~Matsui, H.~Murai, S.~Arahira, S.~Kutsuzawa, and Y.~Ogawa, ``{30-GHz}
  bandwidth $1.55~\mu$m strain-compensated {InGaAlAs-InGaAsP} {MQW} laser,''
  {\em IEEE Photonics Technol. Lett.}, vol.~9, no.~1, pp.~25--27, 1997.

\bibitem{kitamura_high-power_1983}
M.~Kitamura, M.~Seki, M.~Yamaguchi, I.~Mito, K.~Kobayashi, and T.~Matsuoka,
  ``High-power single-longitudinal-mode operation of $1.3~\mu$m {DFB-DC-PBH}
  {LD},'' {\em IEEE Elec. Lett.}, vol.~19, no.~20, pp.~840--841, 1983.

\bibitem{ref:Diehl}
L.~Diehl, B.~G. Lee, P.~Behroozi, M.~Loncar, M.~A. Belkin, F.~Capasso,
  T.~Aellen, D.~Hofstetter, M.~Beck, and J.~Faist, ``{Microfluidic tuning of
  distributed feedback quantum cascade lasers},'' {\em Opt. Express}, vol.~14,
  pp.~11660--11667, Nov. 2006.

\bibitem{ref:Moreau_V3}
V.~Moreau, R.~Colombelli, R.~Perahia, O.~Painter, L.~R. Wilson, and A.~B.
  Krysa, ``{Proof-of-principle of surface detection with air-guided quntum
  cascade lasers},'' {\em Opt. Express}, vol.~16, pp.~8387--6396, Apr. 2008.

\bibitem{ref:Huang_MCY}
M.~C.~Y. Huang, Y.~Zhou, and C.~J. {Chang-Hasnain}, ``A nanoelectromechanical
  tunable laser,'' {\em Nature Photon.}, vol.~2, pp.~180--184, Mar. 2008.

\bibitem{ref:Lee_MCM}
M.-C.~M. Lee and M.~C. Wu, ``{MEMS-Actuated microdisk resonators with variable
  power coupling ratios},'' {\em IEEE Photonics Tech. Lett.}, vol.~17,
  pp.~1034--1036, May 2005.

\bibitem{ref:Bowen_WP}
K.~H. Lee, T.~G. McRae, G.~I. Harris, J.~Knittel, and W.~P. Bowen, ``{Cooling
  and control of a cavity opto-electromechanical system},'' {\em
  arXiv:0909.0082v3 [quant-ph]}, Sept. 2009.

\bibitem{kippenberg_cavity_2008}
T.~J. Kippenberg and K.~J. Vahala, ``Cavity optomechanics: {Back-Action} at the
  mesoscale,'' {\em Science}, vol.~321, pp.~1172--1176, Aug. 2008.

\bibitem{chan_optical_2009}
J.~Chan, M.~Eichenfield, R.~Camacho, and O.~Painter, ``Optical and mechanical
  design of a ``zipper" photonic crystaloptomechanical cavity,'' {\em Opt.
  Express}, vol.~17, pp.~3802--3817, Mar. 2009.

\bibitem{eichenfield_picogram-_2009}
M.~Eichenfield, R.~Camacho, J.~Chan, K.~J. Vahala, and O.~Painter, ``A
  picogram- and nanometre-scale photonic-crystal optomechanical cavity,'' {\em
  Nature}, vol.~459, pp.~550--555, May 2009.

\bibitem{camacho_characterization_2009}
R.~M. Camacho, J.~Chan, M.~Eichenfield, and O.~Painter, ``Characterization of
  radiation pressure and thermal effects in a nanoscale optomechanical
  cavity,'' {\em Opt. Express}, vol.~17, no.~18, pp.~15726--15735, 2009.

\bibitem{johnson_perturbation_2002}
S.~G. Johnson, M.~Ibanescu, M.~A. Skorobogatiy, O.~Weisberg, J.~D.
  Joannopoulos, and Y.~Fink, ``Perturbation theory for maxwell's equations with
  shifting material boundaries,'' {\em Phys. Rev. E}, vol.~65, p.~066611, June
  2002.

\bibitem{eichenfield_optomechanical_2009}
M.~Eichenfield, J.~Chan, R.~M. Camacho, K.~J. Vahala, and O.~Painter,
  ``Optomechanical crystals,'' {\em Nature}, vol.~462, pp.~78--82, Nov. 2009.

\bibitem{MEMS_tas_stiction_1996}
N.~Tas, T.~Sonnenberg, H.~Jansen, R.~Legtenberg, and M.~Elwenspoek, ``Stiction
  in surface micromachining,'' {\em J. Micromech. Microeng.}, vol.~6, no.~4,
  pp.~385--397, 1996.

\bibitem{MEMS_maboudian_critical_1997}
R.~Maboudian and R.~T. Howe, ``Critical review: Adhesion in surface
  micromechanical structures,'' {\em J. Vac. Sci. Technol., B}, vol.~15, no.~1,
  pp.~1--20, 1997.

\bibitem{rosenberg_static_2009}
J.~Rosenberg, Q.~Lin, and O.~Painter, ``Static and dynamic wavelength routing
  via the gradient optical force,'' {\em Nature Photon.}, vol.~3, no.~8,
  pp.~478--483, 2009.

\bibitem{kippenberg_cavity_2007}
T.~J. Kippenberg and K.~J. Vahala, ``Cavity {Opto-Mechanics},'' {\em Opt.
  Express}, vol.~15, pp.~17172--17205, Dec. 2007.

\bibitem{braginsky_quantum_1995}
V.~B. Braginsky, F.~Y. Khalili, and K.~S. Thorne, {\em Quantum Measurement}.
\newblock Cambridge University Press, May 1995.

\bibitem{lin_mechanical_2009}
Q.~Lin, J.~Rosenberg, X.~Jiang, K.~J. Vahala, and O.~Painter, ``Mechanical
  oscillation and cooling actuated by the optical gradient force,'' {\em Phys.
  Rev. Lett.}, vol.~103, no.~10, pp.~103601--4, 2009.

\bibitem{ref:Perahia_R}
R.~Perahia, T.~P.~M. Alegre, A.~H. {Safavi-Naeini}, and O.~Painter,
  ``Surface-plasmon mode hybridization in subwavelength microdisk lasers,''
  {\em Appl. Phys. Lett.}, vol.~95, pp.~201114--3, Nov. 2009.

\bibitem{ref:Hill_MT}
M.~T. Hill, Y.~S. Oei, B.~Smalbrugge, Y.~Zhu, T.~D. Vries, P.~J. Veldhoven,
  F.~W. M.~V. Otten, J.~P. Turkiewicz, H.~D. Waardt, E.~J. Geluk, S.~H. Kwon,
  Y.~H. Lee, R.~Notzel, and M.~K. Smit, ``{Lasing in metallic-coated
  nanocavities},'' {\em Nature Photon.}, vol.~1, pp.~589--594, Oct. 2007.

\bibitem{ref:Carmichael_Rice}
P.~R. Rice and H.~J. Carmichael, ``{Photon statistics of a cavity-QED laser: A
  comment on the laser-phase transition analogy},'' {\em Phys. Rev. A},
  vol.~50, pp.~4318--4329, Nov. 1994.

\bibitem{ref:Yamamoto_Bjork4}
G.~Bj$\ddot{\text{o}}$rk, A.~Karlsson, and Y.~Yamamoto, ``{On the line width of
  lasers},'' {\em Appl. Phys. Lett.}, vol.~60, pp.~304--306, Jan. 1992.

\bibitem{slusher_optical_1994}
R.~E. Slusher, ``Optical processes in microcavities,'' {\em Semicond. Sci.
  Technol.}, vol.~9, no.~{11S}, pp.~2025--2030, 1994.

\bibitem{loncar_dynamically_2009}
I.~W. Frank, P.~B. Deotare, M.~W. {McCutcheon}, and M.~Loncar, ``Dynamically
  reconfigurable photonic crystal nanobeam cavities,'' {\em 0909.2278}, Sept.
  2009.

\bibitem{tank_sillanpaa_accessing_2009}
M.~A. Sillanpaa, J.~Sarkar, J.~Sulkko, J.~Muhonen, and P.~J. Hakonen,
  ``Accessing nanomechanical resonators via a fast microwave circuit,'' {\em
  Appl. Phys. Lett.}, vol.~95, pp.~011909--3, July 2009.

\bibitem{ref:Liu_G}
G.~T. Liu, A.~Stintz, H.~Li, T.~C. Newell, A.~L. Gray, P.~M. Varangis, K.~J.
  Malloy, and L.~F. Lester, ``{The influence of quantum-well composition on the
  performance of quantum dot lasers using $\text{InAs/InGaAs}$ dots-in-a-well
  (DWELL) structures},'' {\em IEEE J. Quan. Elec.}, vol.~36, pp.~1272--1279,
  Nov. 2000.

\bibitem{ref:Stintz}
A.~Stintz, G.~T. Liu, H.~Li, L.~F. Lester, and K.~J. Malloy, ``{Low-threshold
  current density 1.3-$\mu$m $\text{InAs}$ quantum-dot lasers with the
  dots-in-a-well (DWELL) structure},'' {\em IEEE Photonics Tech. Lett.},
  vol.~12, no.~6, pp.~591--593, 2000.

\bibitem{srinivasan_cavity_2006}
K.~Srinivasan, M.~Borselli, O.~Painter, A.~Stintz, and S.~Krishna, ``Cavity
  \textsc{Q}, mode volume, and lasing threshold in small diameter {AlGaAs}
  microdisks with embedded quantum dots,'' {\em Opt. Express}, vol.~14,
  pp.~1094--1105, Feb. 2006.

\bibitem{ref:Seo_JW}
J.~W. Seo, T.~Koker, S.~Agarwala, and I.~Adesida, ``{Etching characteristics of
  Al$_x$Ga$_{1-x}$As in (NH$_4$)S$_x$ solutions},'' {\em Appl. Phys. Lett.},
  vol.~60, pp.~1114--1116, Mar. 1992.

\bibitem{ref:Hwang2}
W.-Y. Hwang, J.~Baillargeon, S.~N.~G. Chu, P.~F. Sciortino, and A.~Y. Cho,
  ``{$\text{GaInAsP/InP}$ distributed feedback lasres grown directly on grated
  substrates by solid-source molecular beam epitaxy},'' {\em J. Vac. S. Tech.
  B}, vol.~16, pp.~1422--1425, May 1998.

\bibitem{day_broadband_superconduting_microstrip_2003}
P.~K. Day, H.~G. {LeDuc}, B.~A. Mazin, A.~Vayonakis, and J.~Zmuidzinas, ``A
  broadband superconducting detector suitable for use in large arrays,'' {\em
  Nature}, vol.~425, pp.~817--821, Oct. 2003.

\bibitem{haus_mode-locking_2000}
H.~Haus, ``Mode-locking of lasers,'' {\em IEEE J. Sel. Top. Quantum Electron.},
  vol.~6, no.~6, pp.~1173--1185, 2000.

\end{thebibliography}
%\bibliographystyle{./ieeetr}

\end{document}